\documentclass[11pt, a4paper]{article}
\pdfoutput=1 

\usepackage{jheppub} 

\usepackage[T1]{fontenc} 
\newcommand{\bra}[1]{\langle #1 |}
\newcommand{\ket}[1]{|#1 \rangle}
\newcommand{\ex}[1]{\langle x^{#1} \rangle}

\newcommand{\vev}[1]{\langle {#1} \rangle}
\newcommand{\OO}{\mathcal{O}}

\title{ Bootstrapping More QM Systems}

\author{David Berenstein $^\dagger$}
\author{and George Hulsey $^\ddagger$}
\affiliation{Dept. of Physics, University of California, Santa Barbara\\Santa Barbara, CA 93106}

\emailAdd{$^\dagger$ dberens@physics.ucsb.edu}
\emailAdd{$^\ddagger$ hulsey@physics.ucsb.edu}

\abstract{We test the bootstrap approach for determining the spectrum of one dimensional Hamiltonians. In this paper we focus on problems that have a two parameter search space in the bootstrap approach: the double well and a periodic potential associated to the Mathieu equation. For the double well, we compare the energies with contributions from perturbative and  non-perturbative results, finding good agreement. For the periodic potentials, we notice that the bootstrap approach gives the band structure of the periodic potential, but it has trouble finding the quasi-momentum of the system. To make further progress on the dispersion relation of the bands, new techniques are needed.  }

\begin{document} 
\maketitle
\flushbottom

\section{Introduction}
\label{sec:intro}

Recently, the bootstrap method has emerged as an alternative numerical method to solve certain quantum mechanical systems
\cite{han,Bhattacharya:2021btd,Aikawa:2021eai} with high precision. Simulations from our previous work, on exactly soluble systems in one dimension, show rapid (exponential)  convergence to the exact energies \cite{Berenstein:2021dyf} in systems where the bootstrap search space was one dimensional. This is not only for the ground state energy---many excited states can be resolved with good accuracy and high precision.

The bootstrap approach has been very successful recently in solving problems of critical exponents for non-trivial field theories in higher dimensions than two \cite{El-Showk:2012cjh,El-Showk:2014dwa}.
The main idea that can be called the `bootstrap approach' is  that one can have constraints on certain correlations of a quantum system due to symmetry, giving relations between certain quantities.  There are additional unitarity constraints: positivity of certain coefficients,  correlators or operators. The main method for solving the problem via bootstrap is to solve the (functional) relations between correlations in terms of some free paramaters and then check the unitarity or positivity constraints. This solution is truncated to a finite amount of information.
Data that passes the second truncation test is a candidate for the full solution of the original problem and as one 
increases the amount of checks (the size of the truncation), the space of allowed solutions should shrink.
In a certain sense, one can argue that the constraints of recursion and unitarity are already apparent in the construction of unitary representations of Lie algebras, which can be carried out algebraically.

This idea has been applied to large $N$ systems, where additional properties of large $N$ physics, like factorization,  make the system behave more classically in gauge invariant variables. Behavior like this can be applied to further simplify the problem. The origins of some of these ideas can be traced back to works 
\cite{Jevicki:1982jj,Jevicki:1983wu,Rodrigues:1985aq}, where a collective coordinate formulation of the theory
is turned into an effective potential whose solutions produce the physics of interest. 
More recently, the bootstrap idea has been applied to study QCD \cite{Anderson:2016rcw}, 
matrix models \cite{Lin:2020mme,Kazakov:2021lel} and the collective coordinate method has also been revisited \cite{Koch:2021yeb}. 

We are particularly interested in a simple approach described in \cite{han} to solve quantum mechanical problems using this approach. This basic bootstrap for quantum mechanical systems goes as follows. One assumes that one has a normalizable eigenstate $\ket E$ of the
energy operator. One assumes that there is a list of (not necessarily hermitian) operators $\OO_i$ which have well defined expectation values on $\ket E$ (they are sufficiently bounded) and that they are linearly independent.
Then the following are true:
\begin{eqnarray}
    \bra E \OO^\dagger\OO \ket E & \geq& 0, \label{eq:positive}\\
    \bra E H \OO \ket E = \bra E  \OO H  \ket E &=& E\bra E  \OO \ket E
\end{eqnarray}
 where $\OO=\sum a_i \OO_i$ (in the top equation, we expect that the expectation values of the list of operators that appear in the squares also belong to the linear  span of the $\OO_i$). In particular, it follows that  $\bra E [H, \OO] \ket E=0$.
 Ideally, the list of $\OO_i$ is such that $[H, \OO_i]$, for sufficiently large $i$, can be written as a linear combination of operators involving other $\OO_j$, for $j<i$, so that one has a recursive way to evaluate the expectation values of the $\OO_i$, given some initial data for the recursion. If these conditions hold, we can say that the Hamiltonian in question is amenable to the bootstrap method.
 
 The bootstrap program consists of using the initial data of the recursion, together with $E$, as a search space to look for consistent solutions of the positive definite constraint \eqref{eq:positive}. One truncates at some dimension the span of the $\OO_i$ and then increases the size of the allowed set of operators $\OO_i$ one by one. The  problem we need to solve if for positive definiteness of a finite Hermitian matrix acting on the vector of the $a_i$.
 If the matrix for $i<i'$ is not positive definite, then the one for $i'$ is not positive definite either. The allowed search space at some size of the matrix is a subset of the allowed search space for smaller matrices (the initial data that passes the positivity test at previous iterations).
 
 This positive matrix constraint can be checked numerically given the initial data. After that the problem  becomes algorithmic: find a list of $\OO_i$ and their recursion relations, and then verify positivity order by order in the
 initial data. Initial data that passes all the tests at some order is considered a candidate for a solution of the spectrum of $E$. If the region of parameter space at some level of iteration is sufficiently small, one can obtain precise bounds on $E$ and the other initial data.

On general grounds, as the amount of data needed to solve a problem increases in dimension, one needs to implement good search strategies for finding solutions to the bootstrap equations. If the bootstrap solutions give rise eventually only to isolated points, one hopes that small isolated islands of allowed parameter space will shrink towards the correct solutions exponentially fast. Once the islands are isolated, grid refinements on the allowed space of solutions in the island can be used to zoom in to the correct answer. We can measure the `elliptical' size of the islands as a function of the iteration depth to understand how the overall data is converging.

Before the islands become isolated, the method itself does not have an \textit{a priori} bias for any solution. Because convergence can be very fast, the small islands might become isolated so quickly that a naive grid can miss them, or if the grids are too fine, one can end up wasting computer resources to find solutions.

In this work, we continue our exploration started in \cite{Berenstein:2021dyf} to study systems with a two dimensional search space problem. These are a particle in the double well potential and a particle in a periodic potential, which is associated with solving the Mathieu equation. Our philosophy is to test the method on systems that can be easily understood in other ways in order to understand how to implement the bootstrap ideas efficiently.We can then check how the bootstrap is performing against semi-analytic methods.

 In this work we partially solve some of the issues of how to choose the parameter search space more efficiently.  We use a semiclassical analysis to bias the search space in the double well potential. Basically, we look near classical results for how the data we search for is related to the classical energy. The idea is that solutions should be somewhat near the classical results, especially in semi-classical regimes. 
 
 However, there can be surprises. For example, the spectrum of the energy might be continuous. This is actually what occurs in the second problem we study: a particle in a periodic potential. The bootstrap does not know \textit{a priori} how to implement the periodicity of the wave function. Said another way, the spectrum of the momentum operator is quantized if the particle lives on a circle. The expectation value of the operator does not know about that quantization, as one superposes states with different momenta to get the correct eigenstate. In the bootstrap program that expectation value is a continuous variable. The output is then the spectrum of the particle on a periodic potential without the quantization of the momentum: one gets the band structure of the periodic potential instead. This same observation is found in the recent work \cite{Aikawa:2021eai}.
 In this case, when islands form they shrink in one direction, but remain extended in another.

For this Mathieu problem, we find a different way to bias the answer to take this into account. This has to do with the structure of the recursion relation and some simple bounds on higher moments that can be implemented at little cost in the analysis. 

The paper is organized as follows: in section \ref{sec:dw}, we present results from bootstrapping the double well potential, including agreement between the data and perturbative/non-perturbative analytical predictions. In section \ref{sec:circle}, we describe the problem of a particle on a circle. We discuss the difficulty of imposing periodic boundary conditions in the bootstrap and the aspects of band structure that we expect to appear in the bootstrap results. We then continue present results from bootstrapping a system with a periodic sinusoidal potential. In the appendix we include some more technical details about the algorithmic implementation of the bootstrap.

{\bf Note:} While this work was being completed, the two works \cite{Bhattacharya:2021btd,Aikawa:2021eai} appeared, which have some overlap with our results. Similar to \cite{Aikawa:2021eai}, we find that in periodic potentials the bootstrap method gives rise to the band structure of the potential, rather than a discrete problem for energy eigenvalues.

\section{The Double Well}\label{sec:dw}
For our treatment of the double well, we take the following Hamiltonian:
\begin{equation*}
    H = p^2 + gx^2 + x^4
\end{equation*}
Note that we use $m = 1/2$ which matches the conventions of \cite{han}. We will consider $g < 0$ which is the regime in which we expect non-perturbative contributions due to instantons. In these conventions, states with $E<0$ live in the double well and states with $E>0$ live above it. Our goal is to identify the performance of the bootstrap method to systems which exhibit non-perturbatively suppressed behavior. The bootstrap approach to this system also was studied in \cite{Bhattacharya:2021btd}, released as we were finishing this work. Our analysis confirms and expands on their results. 

\subsection{Setup and expectations}
The bootstrap is 2-dimensional, in the sense that the values $\{E,\ex{2}\}$ determine all higher moments $\ex{n}$ via the following recursion relation valid for $n \geq 4$:
\begin{equation}\label{eq:dwrec}
    \ex{n} =  \frac{(n-3)}{(n-1)}E\ex{n-4} - \frac{(n-2)}{(n-1)}g\ex{n-2} + \frac{(n-3)(n-4)(n-5)}{4(n-1)}\ex{n-6}
\end{equation}
As usual this reproduces the virial theorem; in the above it occurs at $n = 4$:
\begin{equation}
    E=2g\ex{2} + 3\ex{4}
\end{equation}
There are various aspects of the dynamics that we expect the bootstrap to reproduce. Let us start by considering the system classically. The potential is bounded below by $\min V = -g^2/4$. For any $E > -g^2/4$ the turning points $x_1,x_2$ are defined by $E = V(x_i)$ (care must be taken to choose the correct branches when $E < 0$). The period is defined as an integral over an orbit of the motion:
\begin{equation*}
    T = \oint dt = \int_{x_{1}}^{x_{2}} \frac{1}{\sqrt{E-g x^{2}-x^{4}}} d x
\end{equation*}
Note that with $m = 1/2$ one must be extra careful with prefactors in these equations. The classical analogue of $\ex{2}$ is simply the average of $x(t)^2$ over a period of the motion, which in the integral form is
\begin{equation}\label{eq:clcurve}
    \left\langle x^{2}\right\rangle_{\mathrm{cl}}=\frac{1}{T} \int_{x_{1}}^{x_{2}} \frac{x^{2}}{\sqrt{E-g x^{2}-x^{4}}} d x
\end{equation}
This defines a curve in the $E,\ex{2}$ plane. We expect that quantum states of the system live relatively close to this curve, and that they approach the curve increasingly in the semiclassical limit. This allows an efficient initialization of the search space---instead of searching random points, we can look in a neighborhood of the curve defined by \eqref{eq:clcurve}.

Quantum mechanically, we can make a few predictions about the dependence of various quantities on the coupling $g$. The first of these is the ground state energy. In the limit of deep wells (which is $g \to -\infty$) each well behaves approximately as a harmonic oscillator, and the ground state energy is twofold degenerate. Perturbation theory determines (see e.g. \cite{muller})
\begin{equation}\label{eq:gspert}
    E_{0}(g)=-\frac{g^{2}}{4}+\sqrt{2|g|}-\frac{1}{|g|}-\frac{9}{8 \sqrt{2} |g|^{5 / 2}} + \OO(1/|g|^4)
\end{equation}
The second term gives the harmonic contribution and the latter offer corrections. However, it is well-known that instanton effects break the degeneracy. Performing a one-loop path integral over fluctuations around the instanton solution (see e.g. \cite{kleinert}) we expect to see a ground-state energy splitting
\begin{equation}
    \Delta E_{0}^{(1)}=\frac{8 |g|^{5 / 4}}{2^{-1 / 4}} \frac{1}{\sqrt{\pi}} \exp \left[-\frac{\sqrt{2}}{3} |g|^{3 / 2}\right]
\end{equation}
This contribution is exponentially suppressed in the deep well limit, which is the same regime in which we expect the formula to become increasingly accurate. In fact, we can go a step further and include fluctuation interactions---this is a two-loop calculation in the Feynman jargon. Doing so gives a more refined estimate
\begin{equation}\label{eq:2loop}
    \Delta E_{0}^{(2)}=\frac{8 |g|^{5 / 4}}{2^{-1 / 4}} \frac{1}{\sqrt{\pi}} \exp \left[-\frac{\sqrt{2}}{3} |g|^{3 / 2}-\frac{71 \sqrt{2}}{48 |g|^{3 / 2}}\right]
\end{equation}
In the following sections we present results from bootstrapping this system for a range of couplings $g$. We will see that all the predictions above are supported by the bootstrap data. 

\subsection{Bootstrap implementation}
The basic algorithmic structure is the same as that described in our earlier paper \cite{Berenstein:2021dyf}, with a few technical modifications since the search space is two-dimensional. From a set of points corresponding to candidate values of $E,\ex{2}$, we use the recursion \eqref{eq:dwrec} to generate higher moments up to $\ex{2K-2}$ for some $K$. Then we construct the $K \times K$ Hankel matrix $M_{ij} = \ex{i+j},\ 0 \leq i,j \leq K-1$ and check that it is positive-definite. 

All odd moments vanish by symmetry, so one may be tempted to construct a Hankel matrix from only the even moments: $\tilde{M}_{ij} = \ex{2(i+j)}$. This will require the same positivity conditions but is a weaker constraint. More constraints are furnished by including the vanishing odd moments as matrix elements of $M$.\footnote{This is discussed in more detail in our last paper.} However, both methods converge exponentially; only a subleading speedup is achieved by including all the moments. 

We implemented the bootstrap in \textit{Python}. For a given value of $g$ a cursory search was made in a wide envelope surrounding the classical curve \eqref{eq:clcurve} at some low constraint depth $K_0$. From there the algorithm continued to higher depth, generating new grids of finer resolution around sets of allowed values at each iteration. A more detailed description of the technical implementation is relegated the appendix. We were able to obtain data for values of the coupling $g \in [-8.75,-3]$, and for energies $E \in [-g^2/4,5]$ at each value of $g$. We searched depths $K_0 = 5 \leq K \leq 20$ which was more than sufficient to uncover the non-perturbative behavior. Because of the relatively modest size of the matrices involved, the \textit{float64} datatype used was sufficiently precise to avoid egregious numerical errors. 

\subsection{Bootstrap results}
\begin{figure}[h!]
    \centering
    \includegraphics[scale = 0.55]{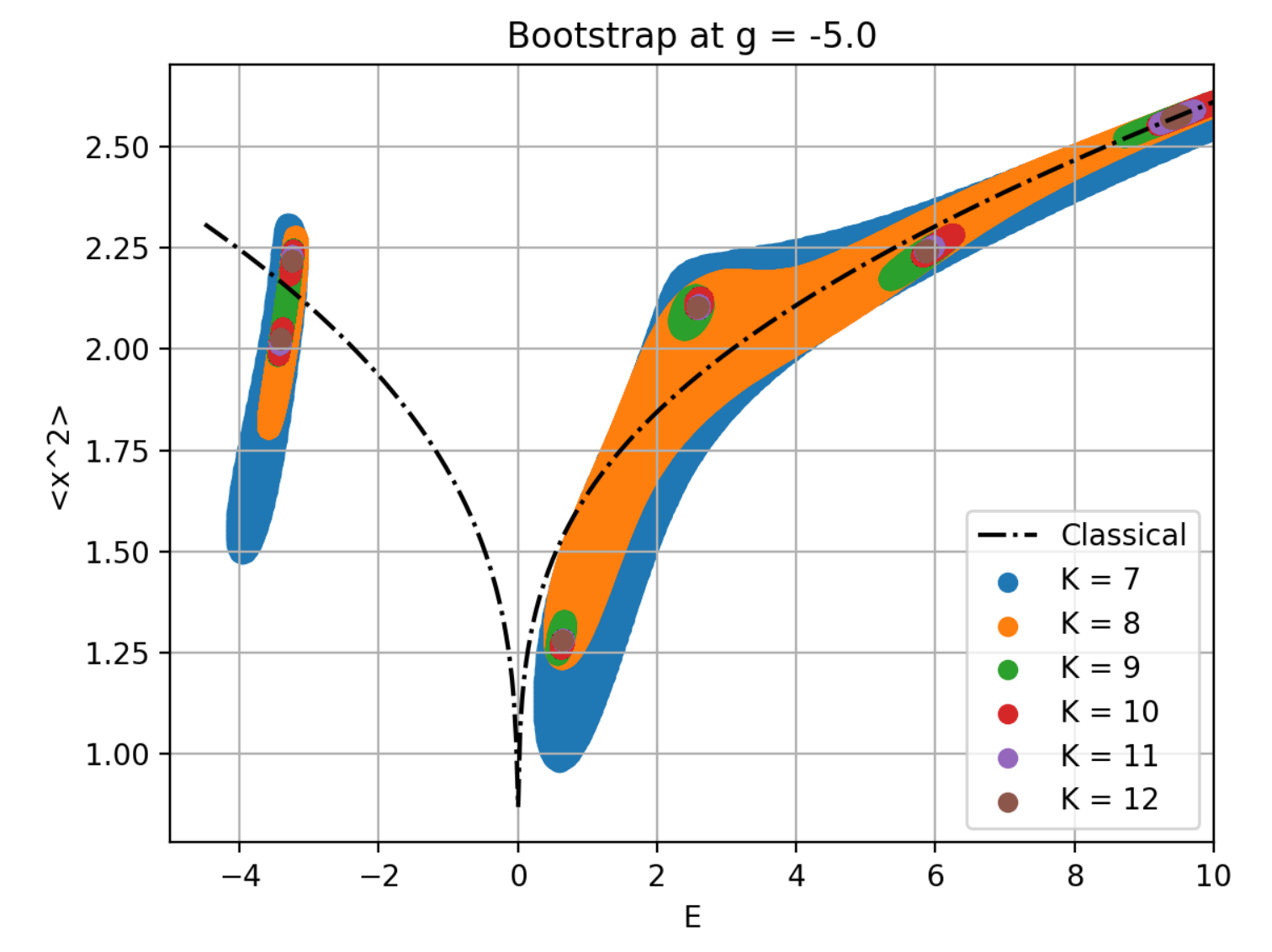}
    \caption{Example bootstrap data for $g = -5$. The ground state splitting is clearly visible as is adherence to the classical curve in the semiclassical regime. Islands form and separate as $K$ increases.}
    \label{fig:example}
\end{figure}

\begin{figure}[h]
    \centering
    \includegraphics[scale = 0.5]{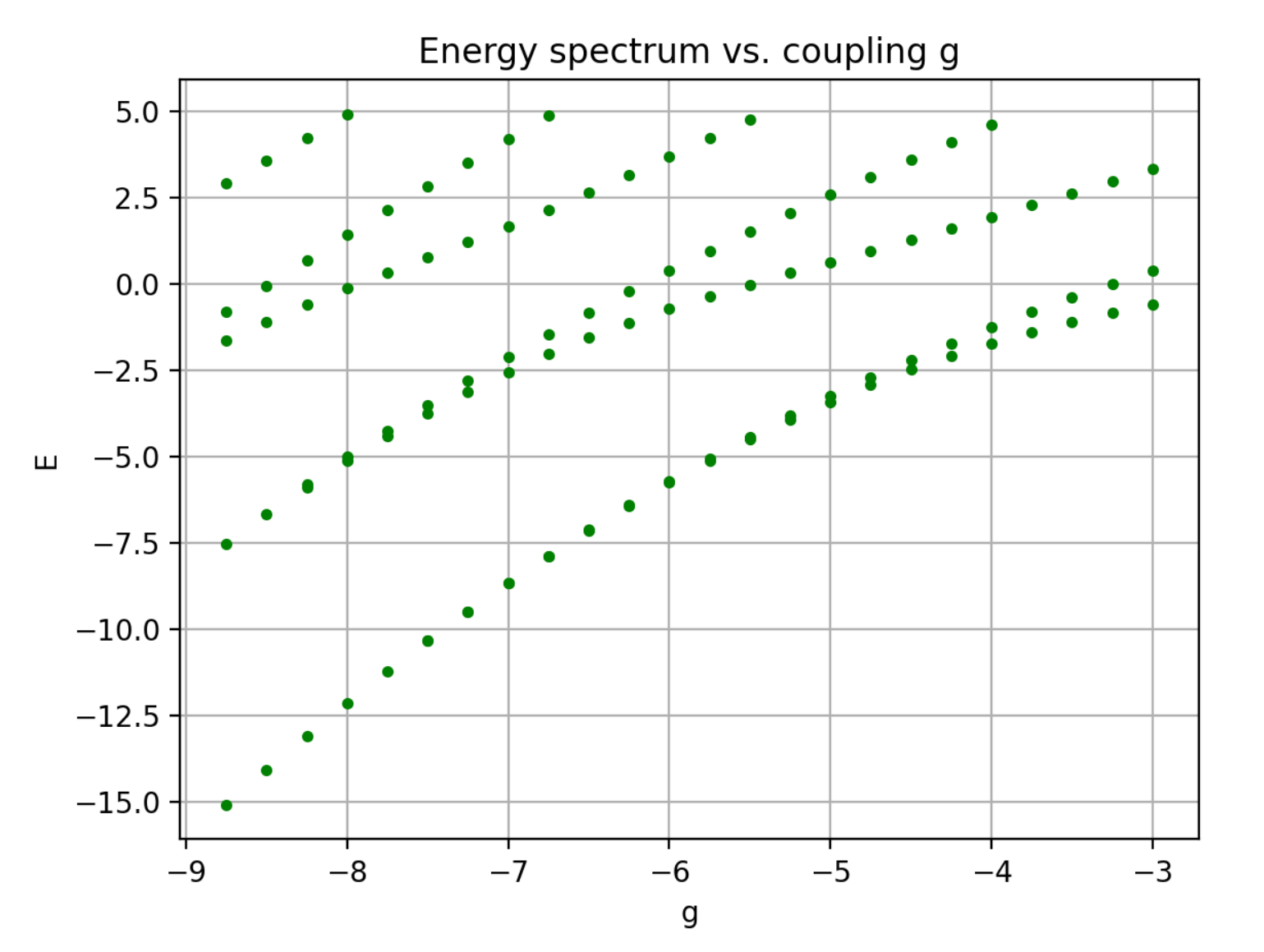}
    \caption{Spectrum versus coupling $g$. The splitting is visible for some values of $g$; for sufficiently low values depths $K > 20$ would be required to resolve the difference. Error bars are included but too small to see.}
    \label{fig:spectrum}
\end{figure}

\begin{figure}[h]
    \centering
    \includegraphics[scale = 0.45]{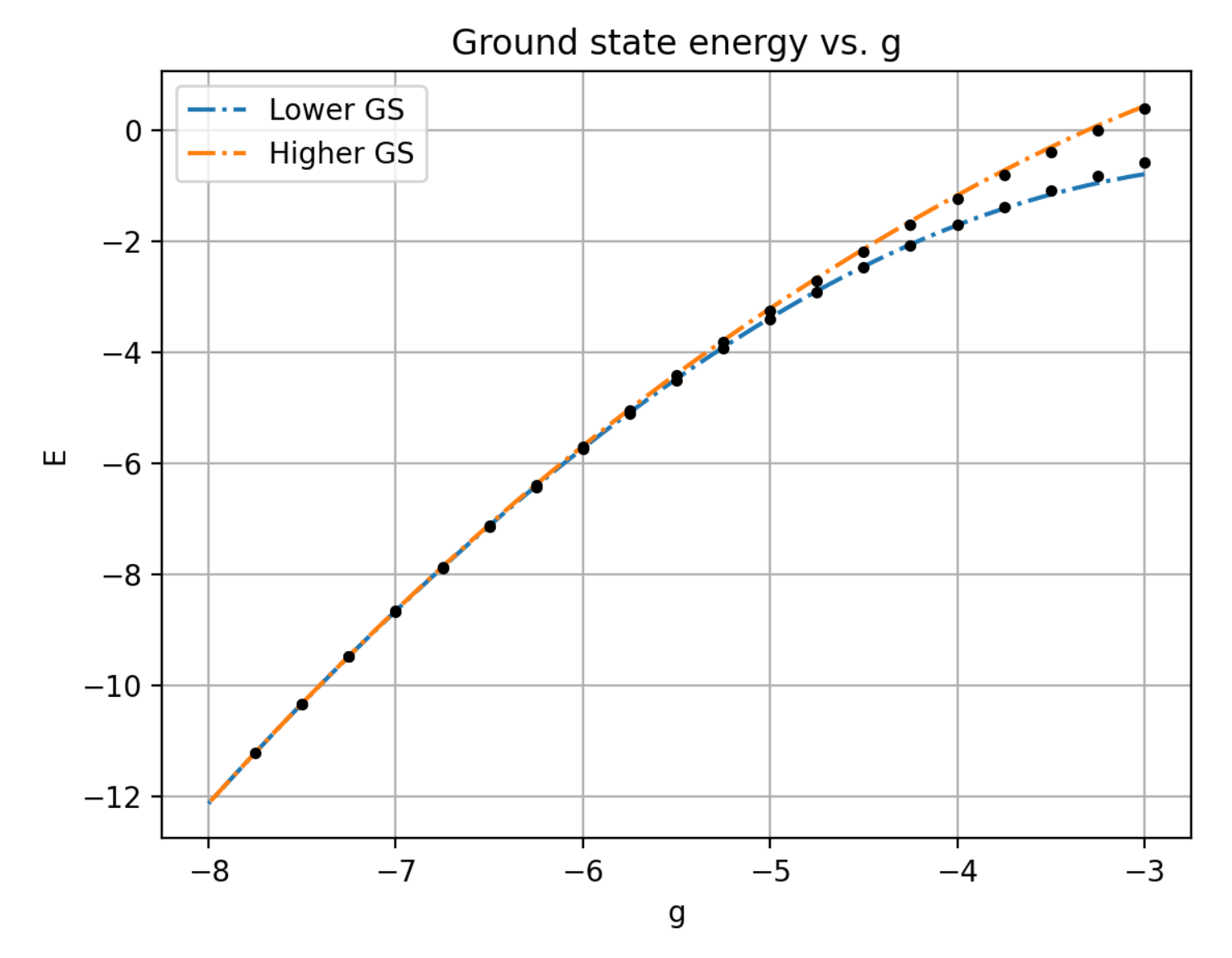}
    \includegraphics[scale = 0.45]{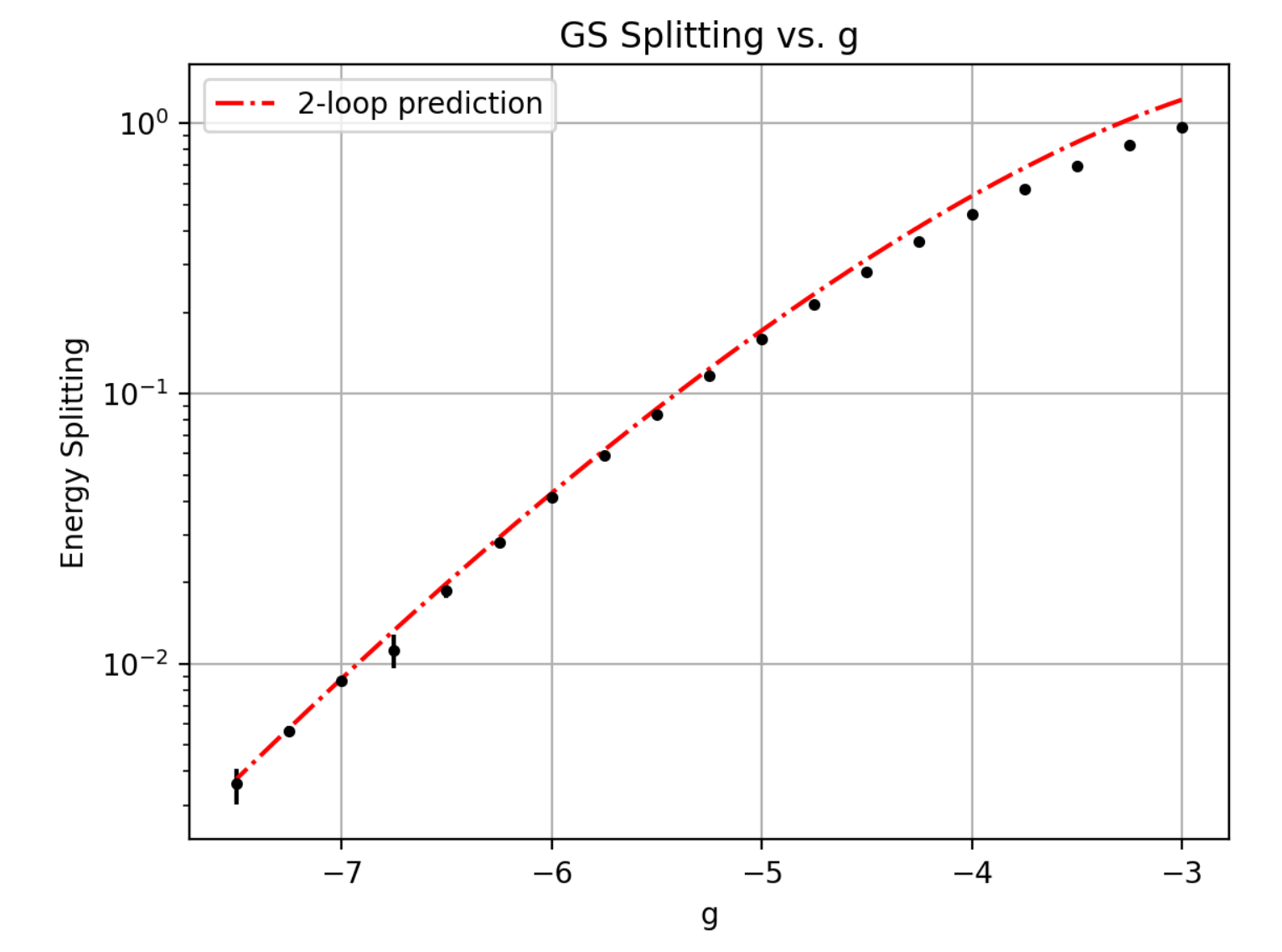}
    \caption{Comparison of non-perturbative formulae to bootstrap data: left, dependence of (split) ground state on $g$; right, dependence of ground state splitting $\Delta E_0$ on $g$. Bootstrap data are black points, predictions are dotted lines. Note the second plot is on a logarithmic scale.}
    \label{fig:gse}
\end{figure}

Here we present some example data from the bootstrap as well as spectral predictions extracted from the bootstrap data. Fig. \ref{fig:example} gives an example of the data from the bootstrap, with the classical curve \eqref{eq:clcurve} superimposed.
\\
The spectral data was obtained by choosing the highest value of $K$ with well-conditioned data for a given coupling. The disparate islands of allowed parameter values were clustered and their energies extracted from the centroids of the clusters. Fig. \ref{fig:spectrum} shows the spectrum, as a function of $g$, extracted in this way. 
With these data we can test the path integral predictions for the splitting. First, we can isolate the ground state energy as a function of $g$. By combining the predictions \eqref{eq:gspert} and \eqref{eq:2loop} we can get a ground state energy estimate which incorporates the splitting. This is included on the left in Fig. \ref{fig:gse}. We can also test the splitting prediction \eqref{eq:2loop} alone, on the right. Agreement is best in the asymptotic regime where this analysis is valid, for large negative $g$. 

\begin{figure}[h!]
    \centering
    \includegraphics[scale = 0.5]{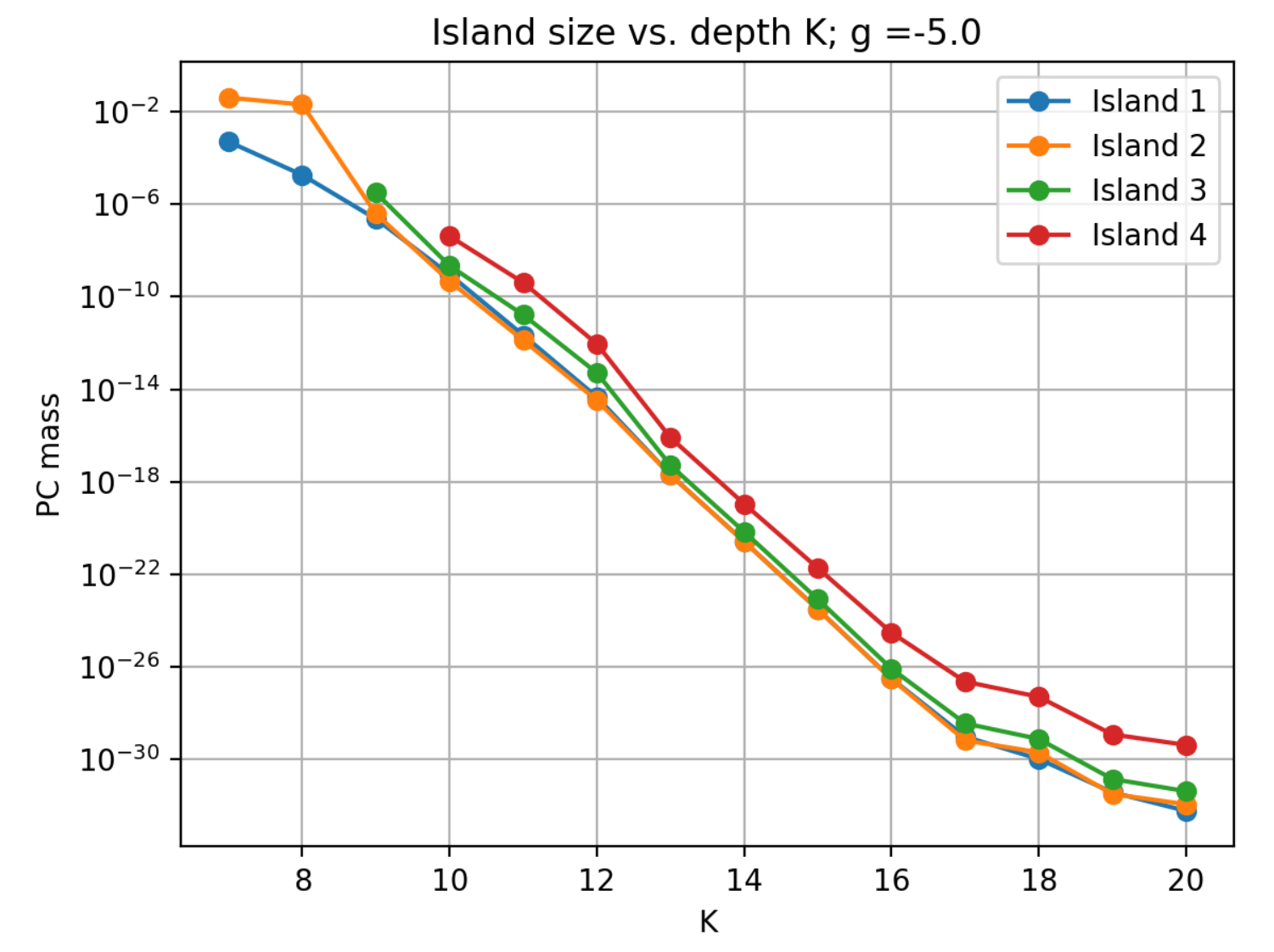}
    \caption{Convergence characterized by `PC mass' (the determinant of the covariance matrix), a rough measure of the elliptical size of the islands in the search space of Fig. \ref{fig:example}. The scale is log-linear.}
    \label{fig:dwconv}
\end{figure}

Finally, we can characterize the convergence properties of the algorithm. To do so we use a rough measure of island size we called `principal component mass' (``PC mass'' in e.g. Fig. \ref{fig:dwconv}). This simply the determinant of the covariance matrix of the set of points which comprise each island. This quantity is proportional to the area of an ellipse with minor and major axes given by the principal components of the point distribution. Fig. \ref{fig:dwconv} depicts the principal component mass for some of the isolated islands shown earlier in Fig. \ref{fig:example}. The convergence is clearly exponential in the constraint depth $K$.

\section{Particle on a circle}\label{sec:circle}
Let us start with a free particle on a circle $\theta\equiv \theta+2\pi$. The free Hamiltonian is given by
\begin{equation}
H= -\frac 12 \partial_\theta^2+V(\theta)
\end{equation}
We are interested in understanding how the bootstrap approach leads to properties of the spectrum of the particle on the circle for arbitrary $V$.
Suppose we have an eigenstate of $H$. To this state, we associate a measure 
\begin{equation}
\mu = |\psi(\theta)|^2 d\theta
\end{equation}  
Normalization of the measure implies that $\langle 1\rangle_\mu =1$. In what follows, we will assume that the expectation values are with respect to $\mu$, so we can safely remove the 
$\mu$ from the expectation values. 
We also have that $\mu\geq 0$ is positive. This can be thought of as a unitarity constraint: probabilities for intervals are positive and normalized.

Consider a complex, differentiable periodic function $f(\theta)$ as a (normal) operator in the quantum system. The general quantum mechanical bootstrap is that 
$\langle  \OO^\dagger \OO \rangle\geq 0$ for all operators. In particular, this applies to $f$. Because of the periodicity, the operator admits a Fourier expansion, 
\begin{equation}
f(\theta) = \sum_{n=-\infty}^\infty a_n \exp(in \theta) 
\end{equation}
Given $\mu$, it follows that
\begin{equation}
\int |f(\theta)|^2 d\mu \geq 0
\end{equation}
Expanding, we get that
\begin{equation}
\int \sum_{n,m} a_m^* a_n\exp(i(n-m)\theta) d\mu \geq 0
\end{equation}
for all $L^2$ normalizable sequences. In particular this is true for all truncations to a finite set of the $a_n$, $-K\leq n \leq \tilde K$. 
We get this way a quadratic form of size $(K+\tilde K+1)\times (K+\tilde K+1)$ which is both a Hermitian matrix and is also positive definite. This form is given by
\begin{equation}\label{eq:toep}
T= \begin{pmatrix}
1 &\langle \exp(i\theta)\rangle &\langle \exp(2 i\theta)\rangle&\dots\\
\langle \exp(-i\theta)\rangle& 1& \langle \exp(i\theta)\rangle & \ddots\\
\langle \exp(-2 i\theta)\rangle &\langle \exp(-i\theta)\rangle& 1& \ddots\\
\vdots & \ddots & \ddots & \ddots 
\end{pmatrix}
\end{equation}
We also have that $T\succeq 0$. This is a special case of a Toeplitz matrix. This positivity condition is a classical result referred to as the trigonometric moment problem.\footnote{Like the Hamburger or Stieltjes moment problem, positivity of this matrix for all ranks guarantees that the sequence $\{\langle e^{in\theta}\rangle\}$ is the moment sequence of a unique positive measure.} Its sufficiency was first proven in 1911 by Toeplitz and Carath\'{e}odory \cite{cara}.

This has consequences. For example, by looking at any $2\times 2$ block with two entries on the diagonal, we find that
\begin{equation}
\begin{pmatrix}
1 &\langle \exp(i(n-m) \theta)\rangle\\
\langle \exp(i(m-n) \theta)\rangle & 1
\end{pmatrix}\succeq 0
\end{equation}   
This produces the {\em obvious} inequality
\begin{equation}
|\langle \exp(i(n-m) \theta)\rangle|^2 \leq 1
\end{equation}
If we specialize to even potentials, we can assume that $\psi(\theta)=\pm  \psi(-\theta)$ so that odd functions in $\theta$ have vanishing expectation value. 
Under those conditions, we have that we can replace $\langle \exp(i(n-m) \theta)\rangle\to \langle \cos((n-m) \theta)\rangle$ and the Toeplitz matrix becomes real symmetric.

Actually, there is a modification we can make to the standard Toeplitz positivity condition, one which improves the convergence of the bootstrap. Let $\OO = \sum a_n e^{in\theta}$ as before. Since the operators $1 \pm \cos(\theta)$ are positive semidefinite on the circle, the following operators are well-defined and unique:
\begin{equation*}
    \OO_\pm = \sqrt{1 \pm \cos(\theta)}\sum_n a_n e^{in\theta}
\end{equation*}
These operators are related to the operator $\OO$ as
\begin{equation*}
    \OO^\dagger\OO = \frac{1}{2}\left(\OO_+^\dagger\OO_+ + \OO_-^\dagger\OO_-\right)
\end{equation*}
On general grounds, we should still have $\langle \OO_\pm^\dagger \OO_\pm\rangle \geq 0$ in any state. Each of these defines a positivity condition on a matrix whose entries are slightly altered from that of the standard Toeplitz matrix \eqref{eq:toep}. Specifically, their matrix elements are
\begin{equation}\label{eq:difftoep}
    T_{nm}^{\pm} = \langle e^{i(n-m)\theta}\rangle \pm \frac{1}{2}\langle e^{i(n-m+1)\theta}\rangle \pm \frac{1}{2}\langle e^{i(n-m-1)\theta}\rangle
\end{equation}
Clearly $2T_{nm} = T^+_{nm} + T^-_{nm}$. Positivity of both $T^\pm$ implies positivity of the matrix \eqref{eq:toep} but not necessarily the other way around. Checking positivity of the matrices with elements given by \eqref{eq:difftoep} is in this sense a stricter condition than simply checking positivity of \eqref{eq:toep}. Indeed, this improves the convergence of the bootstrap, as we checked for ourselves. One may heuristically think that the standard Toeplitz matrix construction associates to some initial data a point in the cone of positive semidefinite matrices, while this modified procedure associates endpoints of a line whose midpoint is the standard Toeplitz matrix. Since the cone of positive matrices is convex this line lies fully within the cone. The line is more likely to leave the cone than the point under perturbation of the initial data.

The general bootstrap problem on the circle is then the following: find a recursion relation for the  Fourier modes of the measure $\langle \exp(i(n-m) \theta)\rangle$ at fixed energy $E$, expressed in terms of some finite amount of data (parameters). Solve the recursion relation and check for positivity at various sizes of the matrix $T$, or of both $T^\pm$. As we increase the size of the matrix, our previous results are still valid, so we can only shrink the allowed parameter space. If the parameter space has shrunk enough, we have `solved' the problem to some precision and ideally we obtain a discrete spectrum for $E$.

\subsection{The free particle}

To understand how the bootstrap works, we should first study the free particle on the circle, with Hamiltonian
\begin{equation}
H_0= \frac {p^2} 2
\end{equation} 
If we solve this system, we are supposed to get 
\begin{equation}
E= \frac{n^2}{2}, \label{eq:expected_energy}
\end{equation}
 for $n\in {\mathbb Z}$, corresponding to $2\pi$-periodic wavefunctions. Consider first that on energy eigenstates we have 
\begin{equation}
\langle [H_0, \OO] \rangle =0
\end{equation}
We now take $\OO= \exp(i n \theta)$. We get that
\begin{equation}
[H_0, \exp(i n \theta) ] =  n \exp(i n\theta) p +\frac{n^2}2  \exp(i n \theta) 
\end{equation}
From here 
\begin{equation}
\vev{\exp(i n\theta) p}=  -\frac n 2 \vev{\exp(i n\theta)}
\end{equation}
Similarly, if we use $\OO =  \exp(i n\theta) p$, we get that (after using that $p^2 = 2E$)
\begin{equation}
2En\vev{\exp(i n\theta)}- \frac {n^3 } 4 \vev{\exp(i n \theta))} =0
\end{equation}
As an aside, from $\OO= p$ we find that $\vev{p^2}\geq 0$, so that the bootstrap produces only positive energy.
What is interesting is that this is a direct constraint on the expectation values of the Fourier modes and there is no recursion. There are two types of solutions.

First, there is $\vev{\exp(i n \theta))}=0$. In that case, the Toeplitz matrix is trivially positive.  
Secondly, there are possible solutions where $\vev{\exp(i n\theta)} \neq 0$ so long as 
\begin{equation}
E= \frac{n^2}{8}
\end{equation}

Let us analyze briefly the second one. If we first compare with \eqref{eq:expected_energy}, the quantization seems to be off: it is as if an integer $n$ is allowed to be a half integer also.
We can ask: are we supposed to consider the first type of solutions where $\vev{\exp(i n\theta)}=\delta_{n,0}$ for all non-trivial $n$? If we follow the bootstrap philosophy, the answer should be an unequivocal yes. They correspond to a constant measure. The only constraint is $E\geq 0$.

What is the meaning of these solutions? Clearly, $p\simeq \sqrt {2 E}$ would be continuous. Nowhere in the commutations relations used to generate the consistency conditions is it specified that the spectrum of $p$ is quantized. All we know is that the potential is periodic, but we don't know that the period is $2\pi$. This was only implicit in the choice of mode functions. We could just as well be on a covering of the circle and the bootstrap would not change, except that now additional values of $n$ that are not integers would also be allowed. If we allow for this possibility, what the bootstrap has produced is a continuous value of $p$ and we have the band structure of the particle on a circle. There is a momentum $p$, and a quasimomentum $p\mod(\Pi)$ where $\Pi$ is the minimal momentum in the dual torus of the circle. In this case $\Pi =1$.

Now let us again examine the second type of solutions where for some integer $n$ (and also $-n$)
, $\vev{\exp(i n \theta)}= C  \neq 0$. For al other $|m| \neq n$, we have $\vev{\exp(i m \theta)}= 0$.
The Toeplitz matrix  then has a  positive definite submatrix of the form
\begin{equation}
M= \begin{pmatrix} 1& C & 0& \dots\\
C^* & 1 & C & \ddots\\
0& C^* &1 &\ddots\\
\vdots &\ddots &\ddots &\ddots
\end{pmatrix}
\end{equation}
This system is very similar to a discrete second order difference operator with Dirichlet boundary conditions
\begin{equation}
 \Delta= \begin{pmatrix} 2& 1 & 0& \dots\\
1 & 2 & 1 & \ddots\\
0& 1 &2 &\ddots\\
\ddots &\ddots &\ddots &\ddots
\end{pmatrix}
\end{equation}
The operator $\Delta$ is positive and gapless in the large matrix limit. We can do a change of basis where the $1$ outside the main diagonal acquire random phases, which can be chosen to be constant. 
We have that 
\begin{equation}
M \simeq |C| \Delta + (1- 2|C|) {\bf 1}
\end{equation}
In the infinite size limit we get that because $\Delta $ is gapless,  $(1- 2|C|)\geq 0$. That is, we find a bound $|C|\leq 1/2$. At finite size, $|C|> 1/2 $, it is like having a negative mass squared term on a lattice. We can get positivity of the matrix with bounds that approach $|C_{\max}|\to 1/2$. The tachyon can be stabiliized by finite size effects, where the smallest eigenvalue of $\Delta $ is of order $1/K^2$.

Consider the wave functions $\psi(\theta) \propto \cos(n \theta+\phi)$ where $n$ is an integer or a half integer. One can check that $|\psi|^2$, properly normalized, saturates  the bound $|C|=1/2$ with a specific phase for $C$.
They are allowed states in the quantum system and the bootstrap is consistent that fact: all solutions that pass the bootstrap bounds are allowed physical states.
What can we do now with states where $|C|<1/2$? The interpretation is that the measure $\mu$ we get is a convex combination of two allowed bootstrap solutions for different angles.
This indicates that the bootstrap solutions we have found analytically are compatible with arbitrary density matrices for a system with two levels. A mixed state measure would be a convex combination of two measures.

These depend on three parameters usually. In this case, we only get a two parameter space, so one can not resolve completely the density matrix problem just from the probability measure $\mu$.
The extremal bounds are pure states and the ones in the middle  of the disk can be either mixed states or pure states. The non-trivial solutions to the bootstrap in this case indicate a double degeneracy. They occur at quasimomentum $q=(p\mod \Pi)=0=-q$ and at $q=(p\mod \Pi)=\Pi/2\equiv -q$; the symmetric points in the band occur where both interfering wave functions have the same quasimomentum.
\subsection{The Mathieu problem}
We now turn to applying these ideas to a particle moving in an inverted cosine potential on the circle. The Hamiltonian we use is 
\begin{equation*}
    H = \frac{p^2}{2} + V(\theta) \equiv \frac{p^2}{2} - 2a \cos(\theta)
\end{equation*}
The time independent Schr\"odinger equation takes the form
\begin{equation*}
    \left[ -\frac{1}{2}\frac{d^2}{d\theta^2} - 2a \cos(\theta)\right]|\psi\rangle = E |\psi\rangle
\end{equation*}
which is equivalent to the classical Mathieu equation after a change of variables:
\begin{equation*}
    \frac{d^{2} y}{d x^{2}}+(A-2 q \cos 2 x) y=0
\end{equation*}

We would like impose periodic boundary conditions $\psi(\theta) = \psi(\theta+2\pi)$. This will turn the Schr\"odinger equation into a Sturm-Liouville problem and hence quantize the energy $E$. Unfortunately, as discussed previously, the 
naive bootstrap system  does not do that---there are no constraints in the moment recursion enforcing periodic boundary conditions. Instead, we should expect to get a band structure for the potential.

The Fourier coefficients of the periodic wave function satisfy a recursion relation and some exact solutions of the corresponding equation are known (see \cite{10.1093/ptep/ptaa024} and references therein). It is not clear that the more general problems of the band structure are solved analytically.

\subsubsection{Recursion relation between moments}
As described previously,  we are interested in trigonometric moments $t_n = \langle e^{in\theta}\rangle$. Some relevant commutators are listed below.
\begin{gather*}
    [p,e^{in\theta}] = ne^{in\theta};\qquad [p^2,e^{in\theta}] = n^2e^{in\theta} + 2ne^{in\theta}p\\
    [p^2,e^{in\theta}p] = n^2e^{in\theta}p +2ne^{in\theta}p^2;\qquad [e^{im\theta},e^{in\theta}p] = -me^{i(n+m)\theta}
\end{gather*}
In particular these imply, with $V(\theta) = -2\alpha\cos(\theta) = -\alpha (e^{i\theta} + e^{-i\theta})$
\begin{equation*}
    [V(x),e^{in\theta}p] = \alpha e^{i(n+1)\theta} - \alpha e^{i(n-1)\theta}
\end{equation*}

Using $\langle [H,\mathcal{O}]\rangle =0$ and $\langle H\mathcal{O}\rangle = E\langle \mathcal{O} \rangle$ for arbitrary operators $\cal O$ in energy eigenstates, we obtain some relations between mixed expectation values of the operators $e^{in\theta},p$. Recall $t_n \equiv \langle e^{in\theta} \rangle$.
\begin{itemize}
    \item $\mathcal{O} = e^{in\theta}$ using $\langle [H,\mathcal{O}]\rangle =0$ yields
    \begin{equation}
        0 = \langle[p^2,e^{in\theta}]\rangle = n^2\cdot t_n + 2n\cdot \langle e^{in\theta}p\rangle
    \end{equation}
    \item $\mathcal{O} = e^{in\theta}p$ using $\langle [H,\mathcal{O}]\rangle =0$ yields
    \begin{equation*}
        0 = \frac{n^2}{2} \langle e^{in\theta}p\rangle + n\langle e^{in
    \theta}p^2\rangle + \alpha t_{n+1} - \alpha t_{n-1}
    \end{equation*}
    \item Finally $\langle \mathcal{O}H\rangle = E\langle \mathcal{O}
    \rangle$ using $\mathcal{O} = e^{in\theta}$ yields
    \begin{equation*}
        \frac{1}{2}\langle e^{in\theta}p^2\rangle - \alpha t_{n+1} - \alpha t_{n-1} = E\cdot t_n
    \end{equation*}
\end{itemize}
Combining the above relations in order to eliminate moments involving the operator $p$ yields the recursion relation for the moments $t_n$:
\begin{equation}
    0 = -\frac{n^3}{4}t_n + 2nEt_n + \alpha(2n+1)t_{n+1} + \alpha(2n-1)t_{n-1}
\end{equation}
This is consistent with our previous analysis at $\alpha=0$.
We can turn this around on its head to obtain an expression for the moment $t_{n+1}$:
\begin{equation}
    t_{n+1} = \frac{1}{\alpha(2n+1)}\left[ \left(\frac{n^3}{4}-2nE\right)t_n - \alpha(2n-1)t_{n-1}\right]
\end{equation}
Notice that since the potential is even, the wavefunctions squared can be chosen to be even. This means the measures $\mu$ from which the moments are extracted are all even. Hence  $\forall n$:
\begin{equation}
    t_n = \langle e^{in\theta}\rangle = \langle \cos(nx)\rangle + i\langle \sin(nx)\rangle = \langle \cos(nx)\rangle
\end{equation}
Some special cases of the recursion relation are:
\begin{itemize}
    \item $n = 0$; gives $t_1 = t_{-1}$, which shows directly that $\vev{\sin(\theta)}=0$
    \item $n = 1$; gives
    \begin{equation*}
        t_2 = \frac{1}{3\alpha}\left[ \left(\frac{1}{4}-2E\right)t_1 - \alpha\right]
    \end{equation*}
    where we are taking $t_0 = 1$. This shows that the search space is two dimensional: $S = \{E, t_1\}$, for example.
\end{itemize}

Moreover, as we argued at the beginning of the section, the bootstrap for $2\times2$ matrices implies that $t_1\in [-1,1]$, and also that $t_n\in [-1,1]$. The first constraint gives a bound on the search space for $t_1$. The latter bound can be used as a simplified bootstrap, where we only check the bounds on the $t_n$ without determining full positivity. This can help narrow the search space. Because $t_n$ ends up being a linear function of $t_1$ at fixed energy, we can obtain bounds on $t_1$ by solving $t_n(E,t_1)=\pm 1$, to obtain bounds $t^\pm_1(E,n)$. The allowed values must be between these two solutions. This can be done more generally in periodic problems, where one would obtain linear constraints on the initial data.

\begin{figure}
    \centering
    \includegraphics[width=10 cm]{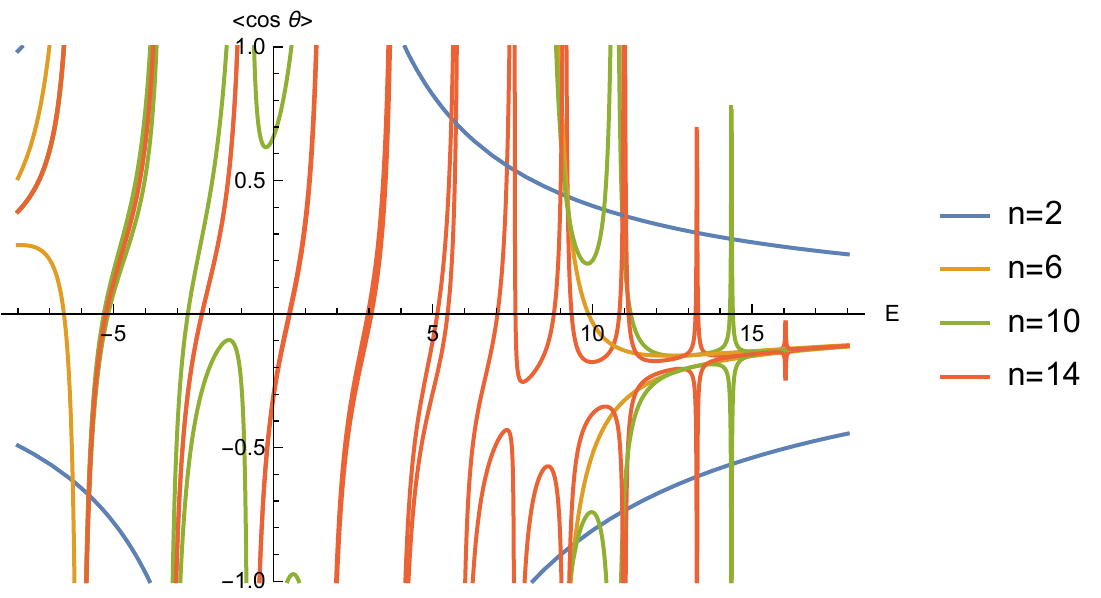}
    \caption{Bounds on $t_1$ from $t_n=\pm 1$ at different values of the energy.} Calculations done with $a=4$. 
    \label{fig:tnbounds}
\end{figure}

Bound on $t_n$ for various $n$ can be seen in Figure \ref{fig:tnbounds}. As can be seen, the allowed values of $t_1=\vev{\cos(\theta)}$ converge rapidly to very shallow strips, more or less nested between each other. Here we are already starting to see the appearance of many bands, and when we take the last value $\vev{\cos(14 \theta)}$ into account, convergence to a narrow strip at high values of $E$. 

In Fig. \ref{fig:tnbounds}, curves are shown that clearly go to un-physical values $t_1\to\pm \infty$. These are poles in the energy denominators that appear when solving for $t_n=\pm1$ for $t_1$. At those values, one is in a gap of energy. At least from this point of view, for this two dimensional problem,  one sees the appearance of gaps in the spectrum relatively easily.

This type of ``simplified bootstrap'' where one checks simple bounds on $t_1$ seems very effective at narrowing the search space more generally and it is interesting to explore further. Passing all tests for the various $t_n$ might even give the band structure exactly, which would be easier than checking numerically for positive definiteness of large matrices.

\subsubsection{Mathieu Bootstrap}
Here we display our results for bootstrapping the potential $V(x) = -2a\cos(x)$ using the trigonometric moment approach and recursion described previously. The algorithmic structure is extremely similar to that of the double well and is additionally elaborated upon in the appendix. We searched a range of values for the potential strength $a$ in the region of the $E,\vev{\cos(x)}$ plane which is expected to contain the bound states: $E \in [-2a,2a]$ and $\vev{\cos(x)} \in [-1,1]$. There is no reason, in principle, why we could not extend the energy range to reach above the maximum of the potential. However, the behavior of the parameter islands changes drastically when $E > 2a$, and so we neglect to pursue this for the time being. 

\begin{figure}[h!]
    \centering
    \includegraphics[scale = 0.7]{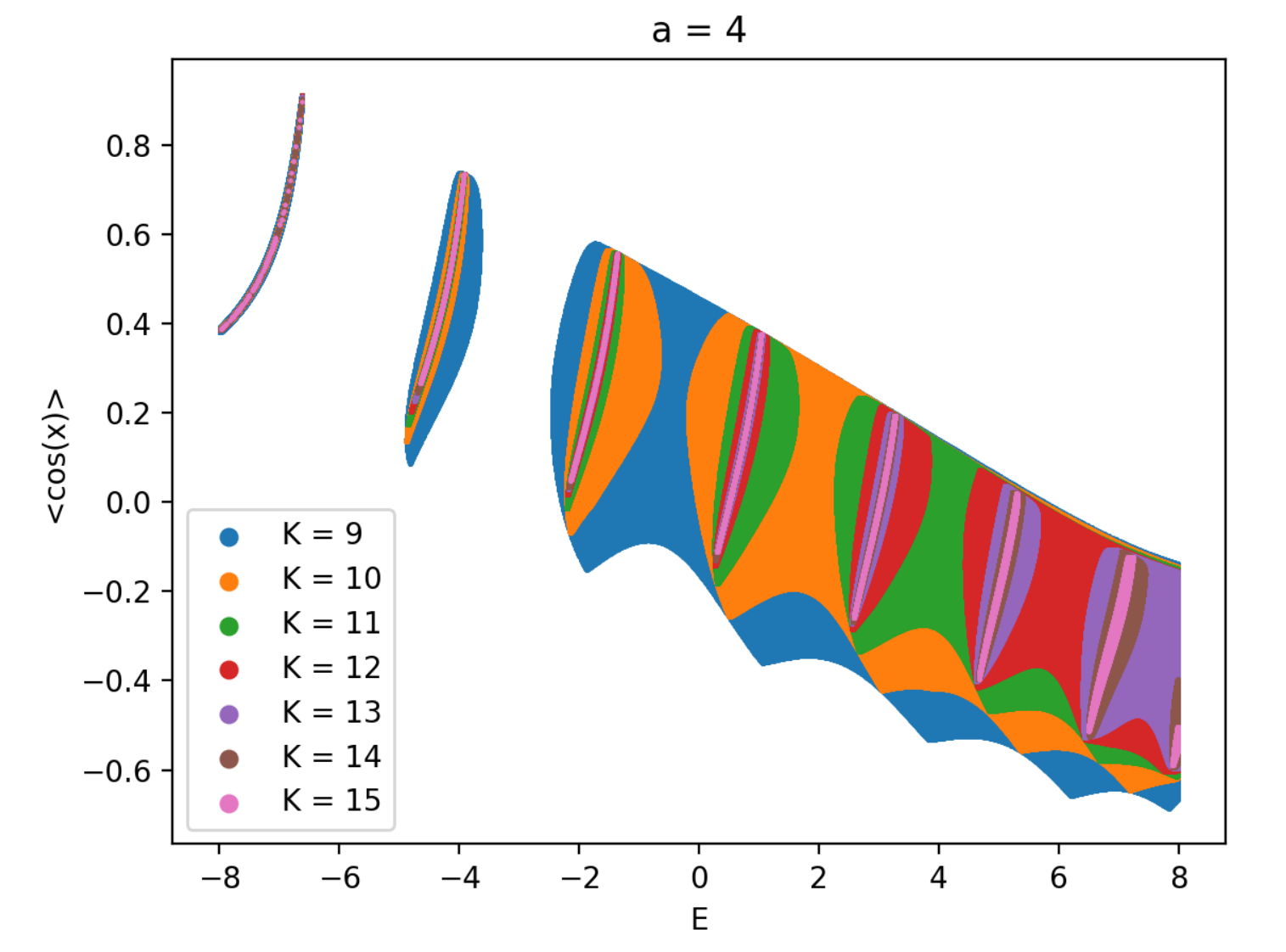}
    \caption{Example data from the Mathieu bootstrap. At high depths, the islands become curves. One can see numerical artifacts due to insufficient resolution appearing in the ground state.}
    \label{fig:mathex}
\end{figure}

\begin{figure}[h!]
    \centering
    \includegraphics[scale = 0.6]{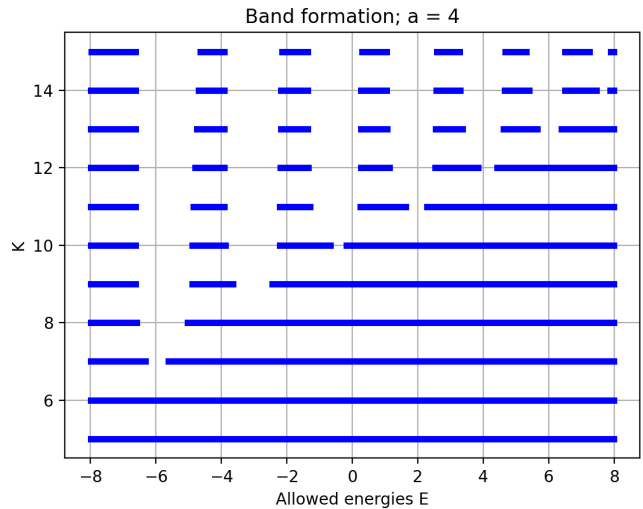}
    \caption{Allowed energy values versus depth $K$. Bands form and converge to limiting values.}
    \label{fig:bands}
\end{figure}

The Fig. \ref{fig:mathex} shows a set of example bootstrap data for $a = 4$. In contrast to the double well problem, the bootstrap converges to curves, rather than points, in the $(E, \langle \cos(x)\rangle)$ plane. 
To get a sense of the convergence, we can extract from the data of Fig. \ref{fig:mathex} the maximum and minimum energy values of each island. These will be upper and lower bounds, respectively, on the exact energy bands. As $K$ increases, one can see these bands form, shrink, and approach a constant width as the islands approach curves of finite length. This is shown in Fig. \ref{fig:bands}. As the bands form they quickly approach a width which persists to higher depth. 
Finally, we can characterize the way the bands change with the potential strength $a$. As $a$ increases, the potential well admits more and more bound states. We can approximate the energies of the low lying states by assuming an approximate harmonic oscillator at the bottom of the potential. 
The squared frequency at the bottom of the well is $\omega^2=2a$. For the case above, the energies of the bands should roughly be given by $E_k\simeq -2 a+ (\frac 12 +k) \omega$. In the case in the figure, these are split by $\omega = 2\sqrt2$ and start at $E_0 \simeq -7$. This is roughly observed. Additional perturbative corrections from the full cosine potential are expected and are negative in first order perturbation theory. Because the bands are somewhat thick, tunneling effects that determine the band structure are important as well.
\begin{figure}[h!]
    \centering
    \includegraphics[scale = 0.6]{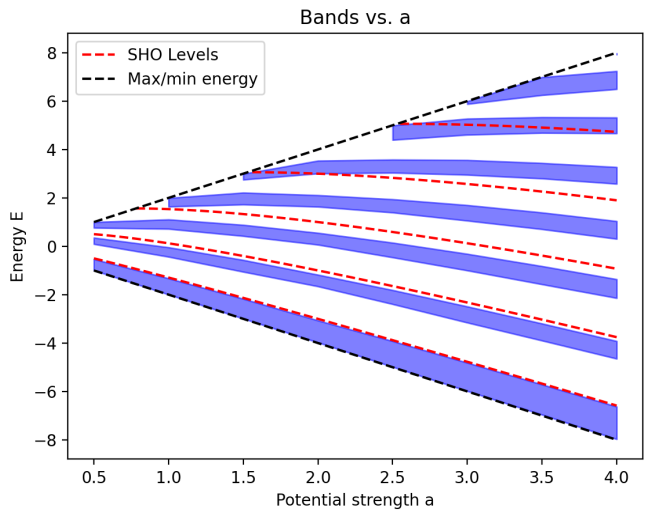}
    \caption{Bands versus potential strength $a$. Bands were selected using the highest value of $K$ with well-conditioned data. This ranges from $K = 8$ for $a = 0.5$ to $K = 14$ for $a \geq 3.5$. Convergence is slower for larger values of $a$. The harmonic approximation is included in red. }
    \label{fig:my_label}
\end{figure}
\subsubsection{Dispersion relations}

\begin{figure}[h!]
    \centering
    \includegraphics[width=6 cm]{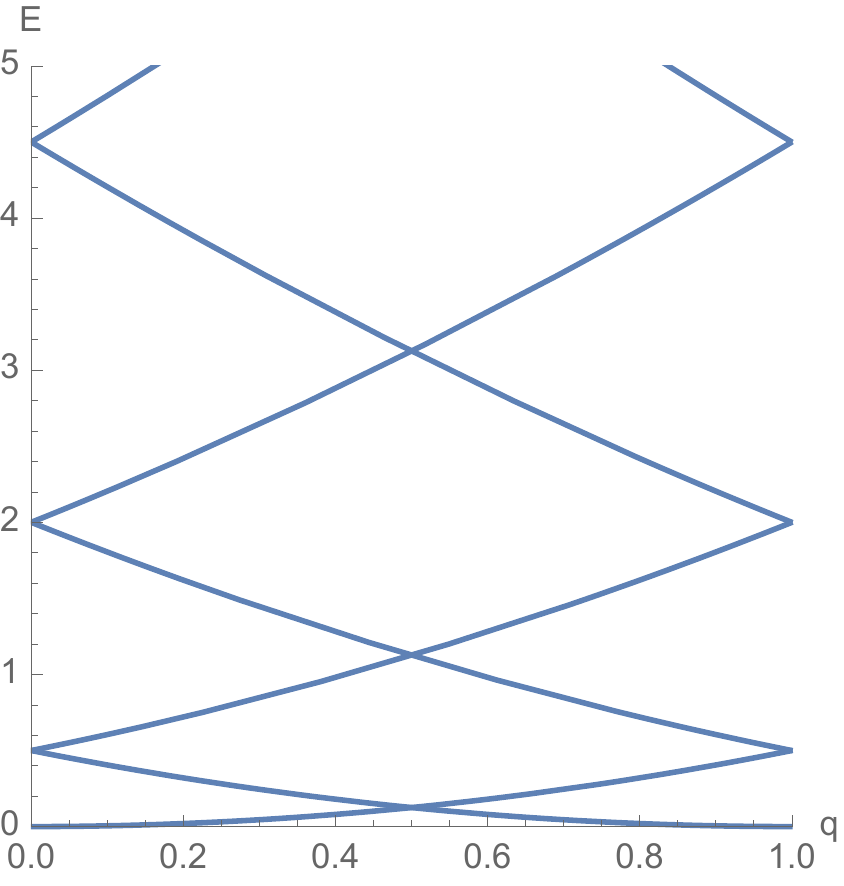}
    \caption{Band structure versus quasimomentum for a free particle}
    \label{fig:disp}
\end{figure}

Our results so far for the Mathieu problem give continuous  segments in the $(E, \vev{\cos(\theta)})$ plane. The general theory of second order differential equations state that there can be at most two linearly independent solutions with the same energy. Because the notion of quasimomentum is conserved, when two states with different quasimomentum have the same energy, it turns out that they have opposite quasimomentum. The only cases where that does not happen is if one is at a symmetry point of quasimomentum: when $q= p \mod(\Pi)\equiv - q$. In figure \ref{fig:disp} we show the band structure for the free particle in the circle.
The bands have (all) the energy values for different quasimomentum. It is exactly when curves interset that there is a double degeneracy, and when our theory results produced non-trivial density matrices. 

Once we add the potential as a perturbation, quasimomentum is conserved, but not momentum. States that are degenerate
in both quasimomentum and energy are usually split by the perturbation and there is level repulsion around those values (the levels, or bands, are usually said to hybridize). This opens up gaps in the spectrum at values of $E\sim n^2/8$, for $n$ integer. The maxima and minima of the bands will end up at symmetric points of the quasimomentum. Hence, we can determine by the bootstrap method the spectrum of the periodic or antiperiodic wavefunctions: we concentrate on the endpoints of the bands. 

We can test this idea for small values of $a$, particularly, considering the pole structure that appears in figures like \ref{fig:tnbounds}. For example, if we take $a=0.1$ as a small parameter, for example, we find that there are poles in the eigth term of the sequence at  $E\simeq 0.494024, 1.12503,2.00051,3.12553,4.50048,6.12541,8.00402$, extremely close to the crossings at $E=n^2/8$, for $n=2, 3,4,5,6,7,8$. This shows that the hybridization  of levels is taking place exactly where it should by perturbation theory arguments.

For other states, determining $q$ is much harder. The values $q, -q$ are degenerate and that fact is  not apparent from
the measure on the circle directly. We known that $\exp(i q) \equiv \exp(i p)$, so in principle we can determine an even function of $q$, $\cos(q)=\cos(p)$. This also works in expectation values.

Unfortunately, since $p$ is unbounded from above as an operator, a series truncation of $\cos(p)$ will have 
large discrepancies at large $p$. Considering that the probability of measuring a given value of $p$ compatible with $q$ is finite, if these are sufficiently suppressed for high values of $p$, one might be able to use this method to obtain and approximate value of $q$. The series of the expectation value of $\cos(p)$ would be obtained from a recursion relation of the expectation values of the even powers of $p$. These expectation values include contributions from high $p$ and are prone to in principle large errors.
This same problem was discovered in \cite{Aikawa:2021eai}.

There is another way to address this problem. We can also study the band structure on a longer spatial period of size $s$ times larger. That would compress the value of $\Pi\to \tilde \Pi=\Pi/s$. There would then be more crossings in the band structure at fixed $a$. If we add a small perturbation on $\cos(\theta/s)$, these new crossings will again hybridize and open small gaps.
The location of these small gaps could then be used to extract the different values of the quasimomentum with respect to $\tilde \Pi$.
 Indeed, the bootstrap at those exact points would allow non trivial expectation values of $\cos(m\theta/s)$ for some integer $m$, exactly like what we had in the free particle case: one should be able to detect the non-trivial density matrix structure at those values. 

The implementations of these computations are beyond the scope of the present work. We are currently studying all of these possibilities: truncation of a series or product formula for $\cos(p)$, or 
hybridizing a more complex set of band structures, or alternatively, checking for allowed non-trivial expectation values. 

\section{Conclusion}

In this work we have studied the bootstrap method for simple one dimensional quantum systems, where one can compare to other techniques for solving the quantum problems. We focused on problems that have a two dimensional search space of parameters. The two problems we focused on are the double well potential and a periodic $\cos(\theta)$ potential.

For the double well potential, we showed that the bootstrap was able to distinguish small non-perturbative tunneling effects and was able to match theoretical predictions for energy splittings. This same observation was made in \cite{Bhattacharya:2021btd}. We supplement those results with convergence plots in the search space, where just like in our previous work \cite{Berenstein:2021dyf}, convergence is exponentially fast. 
Because the islands of allowed values shrink rapidly, one needs algorithms that refine the search effectively. One also needs to have a reasonable hint of how to start looking for solutions to the bootstrap equations in the search space.
We solved this second issue by using semiclassical estimates of where to start looking. For the first problem, our numerical approach is described in detail in the appendices. The codes used to generate this data are available upon request.

For the periodic potential, the bootstrap method gives the band structure of the potential. Maximum and minimum values of each of the bands correspond to either periodic or antiperiodic solutions of the Schr\"odinger problem. The more general problem of determining the dispersion relation of the band as a function of quasimomentum is open. There are technical issues on obtaining the quasimomentum from the bootstrap data to finish solving the band structure  problem that have also been found by Aikawa et al. recently in \cite{Aikawa:2021eai}, while this work was being completed. We are currently investigating ways to address these issues.

The upshot of our results is that the bootstrap technique is very powerful to solve various quantum mechanical problems (at least in none dimension) with exponentially fast convergence, which can be tested against other semianalytic results. In particular, the bootstrap can detect very small energy splittings in problems that might be hard to find in some other way. 
The list of problems accessible to the bootstrap includes problems with a continuous spectrum, where convergence is fast towards the band structure of the problem.

Looking towards the future, we are interested in studying problems in more than one dimension. Particularly problems that can be considered chaotic systems. It will be interesting to see how the bootstrap method performs in these situations, where generically the search space is higher dimensional.

\acknowledgments
We would like to thank R. Brower, S. Catterall, X. Han, O, Janssen, Y. Meurice, J.A. Rodriguez  for discussions and correspondence. Research supported in part by the Department of Energy under grant DE-SC0019139.

\appendix 
\section{Algorithmic Details}\label{app:A}
Here we describe aspects of our algorithmic implementation of the bootstrap for the double well, which should be easily adaptable to similar polynomial potentials. 

The first step, at some fixed value of $g$, is a cursory search at some initial depth (we used $K_0 = 5$). A large Cartesian grid with a moderate resolution was constructed in the energy range $E \in [-g^2/4,5]$ and $\ex{2} \in [0,L]$ where the upper bound $L$ was determined by the curve relating $E$ and $\ex{2}_{\text{cl}}$. This returns some set of allowed values in the $E,\ex{2}$ plane. After this initialization, the algorithm proceeds identically for each successive depth $K_0+1,K_0 + 2,$ etc. The essential details are as follows.
\subsection*{Clustering}
Given some set of allowed values at depth $K$, we first cluster these values into islands. For the double well, this was done using a \verb|KMeans| approach from the scikit-Learn \cite{scikit} toolbox. However, this approach is overkill for spectra without small degeneracy splittings. The reason is that physically we expect some macroscopic energy gap between eigenstates in the absence of non-perturbative splittings. In light of this, an extremely simple option is to project the bootstrap data to the energy axis. Then, one can split groups of points by locating the large gaps between islands corresponding to separated eigenstates. This approach worked well for the Mathieu problem, especially since the `islands' were elongated regions unsuited for most standard clustering algorithms.

To use \verb|KMeans| clustering in the double well, one needs to pass as an argument the number of clusters, then the clustering algorithm proceeds deterministically. To determine the optimal number of clusters we iterated \verb|KMeans| for various numbers of clusters and returned the Davies-Bouldin index \cite{davies}, a diagnostic for clustering success. This index is minimized for well-separated and well-defined clusters; the number of clusters minimizing the index was declared optimal. Given the optimal number of clusters, the allowed values $X_K$ at depth $K$ were clustered via \verb|KMeans| and separated. 
\subsection*{Grid Refinement}
For each island, the algorithm proceeds to generate reformatted grids. For the double well, this was done by finding the smallest rectangle in which the island could be inscribed and generating a new Cartesian grid (for our data, a grid of $1500\times1500$ points) inside the rectangle. This is simple, but works well for islands of uniform shape. 

For the Mathieu problem such an approach is not well-suited. Instead, one can do a principal component analysis. To each separated island is associated a list of points, out of which we may build a $2\times 2$ covariance matrix. By using the eigenvectors of the covariance matrix as primitive basis vectors, one can generate a grid which is automatically resized to account for the shape of the island. In the Mathieu bootstrap we used this approach, scaling the principal axes up so the grid they spanned covered the entire island. For islands with unusual shapes, splitting the island and creating a few such grids is a good strategy.

\subsection*{Bootstrapping}
Given a set of separated islands, the algorithm creates refined grids following one of the procedures outlined above. This defines a new set of trial points for the next depth $K+1$. To increase speed of evaluation, these new points are handled at an array level. We complete the recursion and use a tensor to repackage each moment seqeunce into a Hankel/Toeplitz matrix. We then attempt a Cholesky decomposition (\verb|numpy.linalg.cholesky|) on each matrix. This will fail if the matrix is not positive definite, throwing an error. This can be much faster than explicitly checking the eigenvalues. Then we accept or reject the point, and re-apply the clustering and grid reformatting described above (and move on to higher depths). This constitutes the bootstrap.

\bibliography{refs}

\providecommand{\href}[2]{#2}\begingroup\raggedright\begin{thebibliography}{10}

\bibitem{han}
X.~Han, S.A.~Hartnoll and J.~Kruthoff, \emph{Bootstrapping matrix quantum
  mechanics},
  \href{https://doi.org/10.1103/physrevlett.125.041601}{\emph{Physical Review
  Letters} {\bfseries 125} (2020) }.

\bibitem{Bhattacharya:2021btd}
J.~Bhattacharya, D.~Das, S.K.~Das, A.K.~Jha and M.~Kundu, \emph{{Numerical
  Bootstrap in Quantum Mechanics}},
  \href{https://arxiv.org/abs/2108.11416}{{\ttfamily 2108.11416}}.

\bibitem{Aikawa:2021eai}
Y.~Aikawa, T.~Morita and K.~Yoshimura, \emph{{Application of Bootstrap to
  $\theta$-term}},  \href{https://arxiv.org/abs/2109.02701}{{\ttfamily
  2109.02701}}.

\bibitem{Berenstein:2021dyf}
D.~Berenstein and G.~Hulsey, \emph{{Bootstrapping Simple QM Systems}},
  \href{https://arxiv.org/abs/2108.08757}{{\ttfamily 2108.08757}}.

\bibitem{El-Showk:2012cjh}
S.~El-Showk, M.F.~Paulos, D.~Poland, S.~Rychkov, D.~Simmons-Duffin and
  A.~Vichi, \emph{{Solving the 3D Ising Model with the Conformal Bootstrap}},
  \href{https://doi.org/10.1103/PhysRevD.86.025022}{\emph{Phys. Rev. D}
  {\bfseries 86} (2012) 025022}
  [\href{https://arxiv.org/abs/1203.6064}{{\ttfamily 1203.6064}}].

\bibitem{El-Showk:2014dwa}
S.~El-Showk, M.F.~Paulos, D.~Poland, S.~Rychkov, D.~Simmons-Duffin and
  A.~Vichi, \emph{{Solving the 3d Ising Model with the Conformal Bootstrap II.
  c-Minimization and Precise Critical Exponents}},
  \href{https://doi.org/10.1007/s10955-014-1042-7}{\emph{J. Stat. Phys.}
  {\bfseries 157} (2014) 869}
  [\href{https://arxiv.org/abs/1403.4545}{{\ttfamily 1403.4545}}].

\bibitem{Jevicki:1982jj}
A.~Jevicki, O.~Karim, J.P.~Rodrigues and H.~Levine, \emph{{Loop Space
  Hamiltonians and Numerical Methods for Large $N$ Gauge Theories}},
  \href{https://doi.org/10.1016/0550-3213(83)90180-3}{\emph{Nucl. Phys. B}
  {\bfseries 213} (1983) 169}.

\bibitem{Jevicki:1983wu}
A.~Jevicki, O.~Karim, J.P.~Rodrigues and H.~Levine, \emph{{Loop Space
  Hamiltonians and Numerical Methods for Large $N$ Gauge Theories. 2.}},
  \href{https://doi.org/10.1016/0550-3213(84)90215-3}{\emph{Nucl. Phys. B}
  {\bfseries 230} (1984) 299}.

\bibitem{Rodrigues:1985aq}
J.P.~Rodrigues, \emph{{Numerical Solution of Lattice Schwinger-dyson Equations
  in the Large $N$ Limit}},
  \href{https://doi.org/10.1016/0550-3213(85)90077-X}{\emph{Nucl. Phys. B}
  {\bfseries 260} (1985) 350}.

\bibitem{Anderson:2016rcw}
P.D.~Anderson and M.~Kruczenski, \emph{{Loop Equations and bootstrap methods in
  the lattice}},
  \href{https://doi.org/10.1016/j.nuclphysb.2017.06.009}{\emph{Nucl. Phys. B}
  {\bfseries 921} (2017) 702}
  [\href{https://arxiv.org/abs/1612.08140}{{\ttfamily 1612.08140}}].

\bibitem{Lin:2020mme}
H.W.~Lin, \emph{{Bootstraps to strings: solving random matrix models with
  positivite}}, \href{https://doi.org/10.1007/JHEP06(2020)090}{\emph{JHEP}
  {\bfseries 06} (2020) 090}
  [\href{https://arxiv.org/abs/2002.08387}{{\ttfamily 2002.08387}}].

\bibitem{Kazakov:2021lel}
V.~Kazakov and Z.~Zheng, \emph{{Analytic and Numerical Bootstrap for One-Matrix
  Model and ''Unsolvable'' Two-Matrix Model}},
  \href{https://arxiv.org/abs/2108.04830}{{\ttfamily 2108.04830}}.

\bibitem{Koch:2021yeb}
R.d.M.~Koch, A.~Jevicki, X.~Liu, K.~Mathaba and J.a.P.~Rodrigues, \emph{{Large
  N Optimization for multi-matrix systems}},
  \href{https://arxiv.org/abs/2108.08803}{{\ttfamily 2108.08803}}.

\bibitem{muller}
H.J.W.~M\"uller-Kirsten, \emph{{Introduction to Quantum Mechanics}:
  {Schr\"odinger Equation and Path Integral}}, World Scientific (2012),
  \href{https://doi.org/10.1142/8428}{10.1142/8428}.

\bibitem{kleinert}
H.~Kleinert, \emph{Path Integrals in Quantum Mechanics, Statistics and Polymer
  Physics} (01, 1995), \href{https://doi.org/10.1142/1081}{10.1142/1081}.

\bibitem{cara}
C.~Carath\'{e}odory, \emph{\"{U}ber den variabilit\"{a}tsbereich der
  fourierschen konstanten von positiven harmonischen funktionen}, {\emph{Rend.
  Circ. Mat.} {\bfseries 32} (1911) 193}.

\bibitem{10.1093/ptep/ptaa024}
D.J.~Daniel, \emph{{Exact solutions of Mathieu’s equation}},
  \href{https://doi.org/10.1093/ptep/ptaa024}{\emph{Progress of Theoretical and
  Experimental Physics} {\bfseries 2020} (2020) }
  [\href{https://arxiv.org/abs/https://academic.oup.com/ptep/article-pdf/2020/4/043A01/33114067/ptaa024.pdf}{{\ttfamily
  https://academic.oup.com/ptep/article-pdf/2020/4/043A01/33114067/ptaa024.pdf}}].

\bibitem{scikit}
F.~Pedregosa, G.~Varoquaux, A.~Gramfort, V.~Michel, B.~Thirion, O.~Grisel
  et~al., \emph{Scikit-learn: Machine learning in {P}ython}, {\emph{Journal of
  Machine Learning Research} {\bfseries 12} (2011) 2825}.

\bibitem{davies}
D.L.~Davies and D.W.~Bouldin, \emph{A cluster separation measure},
  \href{https://doi.org/10.1109/TPAMI.1979.4766909}{\emph{IEEE Transactions on
  Pattern Analysis and Machine Intelligence} {\bfseries PAMI-1} (1979) 224}.

\end{thebibliography}\endgroup
\bibliographystyle{JHEP}

\end{document}